# Ionization potentials of crystalline organic thin films: Position dependence due to molecular shape and charge redistribution

Benjamin J. Topham, Manoranjan Kumar and Zoltán G. Soos

Department of Chemistry, Princeton University, Princeton NJ 08544

**Abstract**

In addition to electronic polarization or charge redistribution, the shape of neutral conjugated molecules yields position-dependent ionization potentials and electron affinities in organic thin films. Self-consistent I($n$) and A($n$) are computed in each layer $n$ of 10-layer films of prototypical organics on a metal. The depth dependence of I($n$) is discussed at surfaces of anthracene, $C_{60}$ and PTCDA. The shape contribution can be substantial, up to 0.5 eV, and comes primarily from charge-quadrupole interactions.

## 1. Introduction

Many advances have contributed to the emergence of organic electronics as a promising field with diverse applications. One is the preparation of crystalline thin films. Another is their characterization[1,2] by ultraviolet photoelectron spectroscopy (UPS), x-ray photoelectron spectroscopy (XPS) and inverse photoelectron spectroscopy (IPES). The binding energy (BE) or ionization potential of valence or core holes can now be measured to an accuracy of 100 meV. BE shifts at surfaces[3-6] or at different surfaces[7] have recently been reported. In this paper we draw attention to electrostatic contributions that reflect molecular *shape* and relate previous discussions of charges in organic molecular crystals to thin films. Electrostatics of *neutral* molecules tend to be overlooked, partly because they are absent in atomic crystals or in microelectrostatic models[8,9] and partly because older BE data were not suitable for quantitative comparisons.

Condensed phases are perturbations for noble gases or organic molecules with van der Waals and other weak interactions. Polarization reduces gas-phase ionization potentials $I_G$ and increases electron affinities $A_G$. The defining relations

$$I(s) \quad = \quad I_G - P_+, \quad A(s) \quad = \quad A_G + P_- \qquad (1)$$

include all differences between gas and solid, and I(s) is applicable to valence or core holes, to crystalline or amorphous solids, or to surfaces. Gutmann and Lyons' classic book[10] has comprehensive qualitative discussions of electronic, vibrational and lattice contributions to $P_+$ and $P_-$. Subsequent books[11,9] cover experimental and theoretical studies of $P_\pm$. Electronic polarization is the largest part: $P = (e^2/2a)(1 - 1/\kappa) \sim 1$ eV for a charge in a cavity with radius a = 5 Å in a medium with dielectric constant $\kappa \sim 3$. Theory has focused on self-consistent bulk values.[8,9] The transport gap[12] $E_t = I(s) - A(s)$ for generating ions at infinite separation is also the photoconduction gap,[13] while ion pairs are charge-transfer excited states of crystals. The combination $P_+ + P_-$ appears in $E_t$ or for ion pairs.

Accurate BE data call for $P_+$ at surfaces of crystalline thin films. We developed[14] a self-consistent method for electronic polarization and noted that $P_+$ depends on the macroscopic shape of the sample, while $E_t = P_+ + P_-$ does not. We distinguish between two electronic contributions that both depend on the location of the molecular ion. The larger one is the familiar charge redistribution or polarization near the ion; this is the *entire* P ~ 1 eV in atomic solids or in microelectrostatic models[8,9] that treat molecules as polarizable points. The second contribution to P reflects molecular *shape*. A quadrupole, for example, generates a Coulomb potential $\varphi(\mathbf{r})$ in crystals or films. Positive $\varphi(\mathbf{r}) > 0$ at a molecule destabilizes[13,15] a cation and hence decreases $P_+$ while $\varphi(\mathbf{r}) > 0$ stabilizes an anion and increases $P_-$. The role of molecular shape is developed in Section 2 and shown to act in concert with charge redistribution. Since anisotropy is fixed while redistribution reflects the charge, $P_+$ and $P_-$ are different in general.

The depth dependence of I(s) and A(s) is reported in Section 3 for 10-layer PTCDA films. PTCDA (perylenetetracarboxylicdianhydride, Fig. 2) is a prototypical film former with substantial charge anisotropy due to carbonyl groups and large (~1 eV) differences between $P_+$ and $P_-$. We consider films with the (102) surfaces and other surfaces in which $P_+$ and $P_-$ are reversed. Thin films of $C_{60}$, by contrast, have negligible anisotropy and almost identical $P_+$, $P_-$. Pentacene or anthracene films with *ab* surfaces or rubrene films have intermediate anisotropy, while hetoroatoms in sexithiophene again increase the anisotropy. We consider how molecular shape modulates the BE of holes in different layers of films. Detailed comparisons of electronic polarization of these and other films will be presented separately.[16]

## 2. Molecular shape and charge redistribution

We suppose that an isolated (gas phase) molecule has charge density $\rho^{(0)}(\mathbf{r})$ that includes nuclear charges as $\delta$ functions. The ground states of the cation and anion radicals are $\rho_+^{(0)}(\mathbf{r})$ and $\rho_-^{(0)}(\mathbf{r})$, respectively. We form thin films or arrays $\{a\}$ using bulk crystal structures and consider Coulomb interactions $V_{ab}$ between sites *a* and *b*. We neglect intermolecular overlap, thereby reducing the problem to classical electrostatic interactions between quantum mechanical molecules.[14] As in molecular exciton theory, crystal states are products of molecular functions. Gas-phase charges are fixed sources that polarize the array. The resulting (unknown) $\rho(\mathbf{r}^a)$ depend on the location $a$ of molecules and ions. The total electrostatic energy is formally given by[14]

$$E_{tot}(\{a\}) \;=\; \tfrac{1}{2}\sum_a \int_{V_a} d^3\vec{r}^{\,a}\,\rho^{(0)}(\vec{r}^{\,a})\varphi(\vec{r}^{\,a}) \;=\; \tfrac{1}{2}\sum_a \int_{V_a} d^3\vec{r}^{\,a}\,\rho(\vec{r}^{\,a})\varphi^{(0)}(\vec{r}^{\,a}) \tag{2}$$

where $\varphi(\mathbf{r}^a)$ is the Coulomb potential due to all molecules or ions $b \neq a$. Since the interaction is bilinear in charge, $E_{tot}$ can also be expressed in terms of the potentials $\varphi^{(0)}(\mathbf{r}^a)$ of gas-phase charges. The oriented-gas model[10] has gas-phase quantities throughout and yields the first-order correction $E^{(1)}$ of the array, but it completely neglects electronic polarization. $E_{tot}$ for discrete sources and polarizable points is a general result[17] that has been applied to both atomic[18,19] and molecular crystals.[8,9] An array of neutral nonpolar (centrosymmetric) molecules has finite potentials $\varphi^{(0)}(\mathbf{r}^a)$ due to anisotropic $\rho^{(0)}(\mathbf{r})$. Molecular solids are fundamentally different in this respect from noble-gas solids with isotropic $\rho^{(0)}(\mathbf{r})$ and hence $\varphi^{(0)}(\mathbf{r}^a) = 0$ by Gauss' theorem.

To find I(s) or A(s), we replace a neutral molecule $a = d$ with an ion and define the electrostatic energy as $E_{tot} = E(d)$. The sources $\rho^{(0)}(\mathbf{r})$ are identical except at $a = d$, where the change is $\Delta\rho_d^{(0)} = \rho_+^{(0)}(\mathbf{r}^d) - \rho^{(0)}(\mathbf{r}^d)$ for a cation. The crystal potential is now $\varphi_d(\mathbf{r}^a)$ and $P_+(d) = E_0 - E(d)$ is the difference between two extensive quantities, a neutral lattice and a lattice with a cation at $a = d$. Manipulation of Eq. (2) leads to

$$
\begin{aligned}
P_+(d) \;&=\; E_0 - E(d) \;=\; E_1(d) + E_2(d) \\
E_1(d) \;&=\; -\tfrac{1}{2}\int_{V_d} d^3\vec{r}^{\,d}\,\Delta\rho_d^{(0)}(\vec{r}^{\,d})\varphi_d(\vec{r}^{\,d}) \\
E_2(d) \;&=\; -\tfrac{1}{2}\sum_a \int_{V_a} d^3\vec{r}^{\,a}\,\delta\rho_a(\vec{r}^{\,a})\varphi^{(0)}(\vec{r}^{\,a})
\end{aligned}
\tag{3}
$$

where $\delta\rho_a = \rho_d(\mathbf{r}^a) - \rho(\mathbf{r}^a)$ is the self-consistent change in the array with and without an ion at $a = d$. Although anisotropic $\varphi^{(0)}$ of *neutral* molecules is required for finite $E_2(d)$, molecular shape and charge redistribution are not simply additive and contribute to both $E_1$ and $E_2$. An anion at $a = d$ leads to similar relations for $P_-(d)$. Physically, $E_1$ is positive and represents charge redistribution. $E_2$ has opposite signs for a cation with $\delta\rho_d > 0$ and an anion with $\delta\rho_d < 0$. Given fixed sources $\Delta\rho_d^{(0)}$, one ion has larger $E_1$ when $\Delta\rho_d^{(0)}$ and $\delta\rho_d$ add. The sign of $E_2$ also indicates whether anisotropy and redistribution combine ($E_2$ $> 0$) or oppose ($E_2 < 0$). Anisotropic $\rho^{(0)}(\mathbf{r})$ generates finite potentials that decrease (increase) $P_+$ at molecules where $\varphi^{(0)}$ is positive (negative).

Even without intermolecular overlap, we cannot solve Eq. (2) exactly for $\rho(\mathbf{r})$. That requires the functional derivative $\delta\rho(\mathbf{r})/\delta\varphi(\mathbf{r'})$ for changing the charge density at $\mathbf{r}$ due to a potential change at $\mathbf{r'}$ in the molecule. Quantum cell models of solids or molecules are more tractable.[14] Discrete sites lead to partial derivatives $\partial\rho_m{}^a/\partial\varphi_n{}^a$ for the change of atomic charge at $m$ due to a potential change at atom $n$. Moreover, $\varphi_n{}^a$ is readily included as a site energy at atom $n$ and the resulting electric fields $\mathbf{F}_n{}^a = -\boldsymbol{\nabla}\varphi_n{}^a$ can be computed. The INDO/S scheme[20] is well suited[14,5] for a self-consistent treatment. We also include atomic polarizabilities $\tilde{\alpha}_n$ as corrections to INDO/S charges chosen to match the gas-phase polarizability tensor $\alpha$ of the molecule.[14] Discrete versions of $E_1$ and $E_2$ in Eq. (3) are special cases of Eq. (27) of ref. 14 for any number of ions,

$$E_1(n) \quad = \quad -\tfrac{1}{2}\sum_m \Delta\rho_m^{(0)}\varphi_m^{ion} \tag{4}$$

The sum is over atoms $m$ of the ion in layer $n$. The charges $\Delta\rho_m{}^{(0)}$ are differences between the ion and molecule, and $\varphi_m$ is the self-consistent potential of the film with the ion. The $E_2(n)$ term with $\varphi^{(0)}(\mathbf{r})$ becomes

$$E_2(n) \quad = \quad -\tfrac{1}{2}\sum_a \sum_j \left( \delta\rho_j^a \varphi_j^{a(0)} - \delta\vec{\mu}_j^a \cdot \vec{F}_j^{a(0)} \right) \tag{5}$$

The sum is over atoms $j$ of all molecules for charge redistribution $\delta\rho$ and changes $\delta\boldsymbol{\mu}$ of induced atomic dipoles from the self-consistent neutral film. All potentials and fields in $E_1(n)$ and $E_2(n)$ are based on INDO/S atomic charges and $\tilde{\alpha}$ inputs.

Although no longer self-consistent, we obtain first-order corrections[15] to $E_1(n)$ and $E_2(n)$ for the best available quantum chemical $\rho^{(0)}(\mathbf{r})$. We used density functional theory (B3LYP) with the reasonably large 6-311++G** basis in the GAUSSIAN 03 program.[21] The resulting potentials $\Phi^{(0)}(\mathbf{r}_j{}^a)$ are evaluated at atoms and replace the gas-phase $\varphi_j{}^{a(0)}$ in $E_2(n)$. The $E_1(n)$ correction is to sum over $\Phi^{(0)}(\mathbf{r}_m{}^n) - \varphi_m{}^{a(0)}$ at the ion. Taken together, the two corrections reduce to charge-quadrupole interactions when molecules are shrunk to polarizable points with fixed quadrupoles.

### 3. Thin film on inert metal

We model a crystalline thin film on an inert metal as sketched in Fig. 1 and discussed previously.[4,5] The metal is a constant potential surface, taken as $\phi = 0$, at distance $h$ from the closest atom. Bulk crystal structures are used. N molecular layers produce N image layers, with $n = 1$ next to the metal and $n = N$ at the surface. Translational invariance within layers is used for a lattice of neutral molecules in Eq. (2) with electrostatic energy $E_0$. A cation in layer $n$ has electrostatic energy $E(n)$ and $P_+(n) = E_0 - E(n)$ in Eq. (3). Image charges contribute to the potential $\varphi(\mathbf{r})$ in the film.

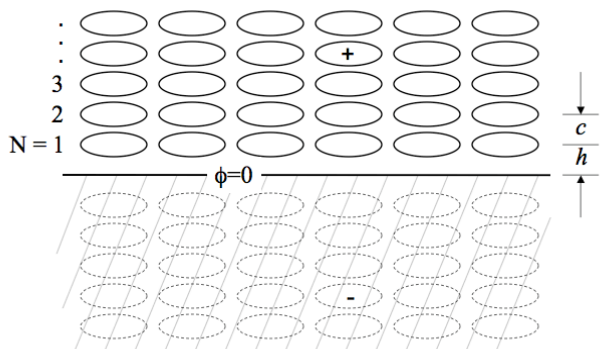

Figure 1. Schematic representation of a film with layer spacing $c$ at separation $h$ from a metal. Dashed ovals are image charge distributions of both molecules and ions.

To illustrate, we consider well-studied PTCDA films[22,4,6] with (102) surfaces, N = 10, $h = 3.2$ Å in Fig. 1 and nearly prone molecules as sketched. The crystal structure is taken from ref. 23. Table 1 lists $E_1(n)$ and $E_2(n)$ for either ion in layer $n$. Fig. 2 shows P($n$) = $E_1(n) + E_2(n)$ with and without B3LYP corrections and average values in layers or bulk. Molecular shape accounts for strikingly different $P_+(n)$ and $P_-(n)$. Large $E_2(n) > 0$ in Table 2 for the cation indicates concerted polarization and anisotropy; $P_+(n)$ depends weakly on $n$ since $E_1(n)$ decreases while $E_2(n)$ increases with $n$. Large $E_2(n) < 0$ for the anion indicates opposed polarization and anisotropy; $P_-(n)$ is small and has strong $n$ dependence since both $E_1(n)$ and $E_2(n)$ increase with $n$. Indeed, $P_-(10) < 0$ at the 102 surface indicates A(s) < $A_G$ due to the crystal potential. The corresponding stabilization of the cation at the surface largely offsets the reduced charge-image interaction and leads to slowly varying $P_+(n)$.

Table 1. Components of P = $E_1 + E_2$ in Eqs. (4) and (5) for a 10-layer PTCDA film with (102) surfaces; $n = 1$ is next to the metal with $h = 3.2$ Å in Fig. 1.

| Layer, $n$ | $E_1$ (eV, cation) | $E_2$ (eV, cation) | $E_1$ (eV, anion) | $E_2$ (eV, anion) |
|:---:|:---:|:---:|:---:|:---:|
| 1 | 1.395 | 0.205 | 0.952 | –0.221 |
| 2 | 1.354 | 0.238 | 0.878 | –0.255 |
| 3 | 1.314 | 0.274 | 0.851 | –0.292 |
| 4 | 1.290 | 0.311 | 0.835 | –0.328 |
| 5 | 1.275 | 0.347 | 0.822 | –0.364 |
| 6 | 1.262 | 0.383 | 0.809 | –0.401 |
| 7 | 1.248 | 0.420 | 0.795 | –0.437 |
| 8 | 1.225 | 0.457 | 0.778 | –0.474 |
| 9 | 1.174 | 0.494 | 0.744 | –0.512 |
| 10 | 0.990 | 0.559 | 0.597 | –0.579 |

We have carried out similar calculations on other films of various thickness for both observed and imagined surfaces.[16] The charge distribution of $C_{60}$ is almost isotropic. The crystal structure from ref. 24 leads as expected to $P_+$ and $P_-$ that differ by less than 10 meV in any layer. The average P is shown in Fig. 3 for a 10-layer film with (1,1,1) surfaces. The BE change between $n = 9$ and 10 is 60 meV. An imagined PTCDA film with (0,0,1) layers in Fig. 3 has partly erect molecules. Two PTCDAs per the unit cell form separate $c/2$ planes. One layer is closest to the metal, the other to the surface. The main points are the reversal of P for the anion and cation, with $P_- > P_+$ in (0,0,1) films, and the large (~1 eV) difference between $P_+$ and $P_-$. Herring bone packing in anthracene crystals[25] leads to $ab$ surfaces with molecular planes that are roughly normal to the surface. Fig. 3 shows $P_+(n)$ and $P_-(n)$ of an anthracene ($ab$) film. Since anthracene and pentacene are alternant hydrocarbons with negligible atomic charges, their shapes and quadrupole moments are responsible for substantial $P_+/P_-$ differences.

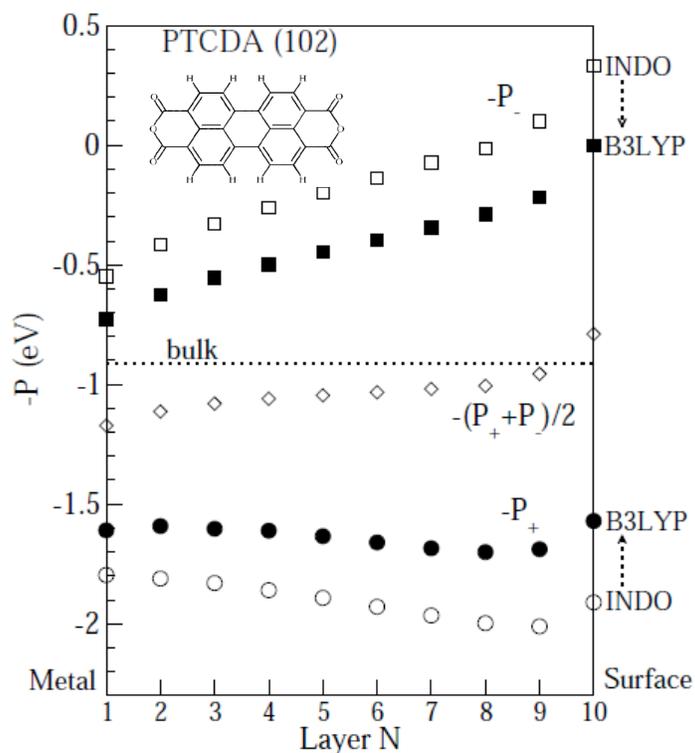

Figure 2. Electronic polarization $P_+$ and $P_-$ from Table 1 of a 10-layer PTCDA film with (102) surfaces. Open symbols refer to INDO/S results, Eqs. (4) and (5); closed symbols refer to B3LYP potentials as discussed in the text.

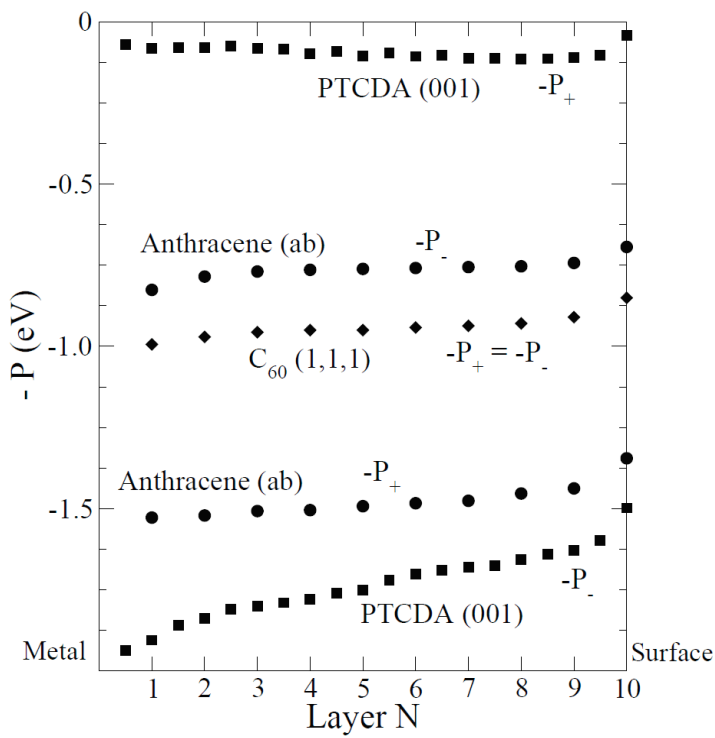

Figure 3. Electronic polarization $P_+$ and $P_-$ of 10-layer films on inert metal: $C_{60}(1,1,1)$ surfaces, anthracene(*ab*) and PTCDA(0,0,1). INDO/S results in Eqs. (4) and (5) are shown with B3LYP potentials as discussed in the text.

Crystalline thin films are two- rather than three-dimensional. The quadrupole potential $1/R^3$ of distant neutral molecules converges as $1/R$ in films. The shape dependence of the bulk can be eliminated theoretically by Ewald constructions[14] that amount to choosing a specific macroscopic shape or experimentally by a common reference. Inversion symmetry at molecules is retained in anthracene crystals or in an *ab* monolayer, but not in *ab* films, although the calculated changes are negligible in the present context. The B3LYP potentials $\Phi^{(0)}(\mathbf{r}_j^a)$ in Fig. 4 at the indicated atoms are quite different in a monolayer, 10-layer film and the bulk. Since H atoms on the periphery are slightly positive, we have $\Phi^{(0)}(\mathbf{r}_j^a) < 0$ in an *ab* monolayer and the cation is stabilized. But H atoms are closest in the next *ab* layers and potentials become less negative. The dashed lines are the average $\Phi^{(0)}$ for the C atoms of the central ring. The bulk value of 170 mV is 360 mV less negative than the monolayer and agrees well with an independent microelectrostatic stabilization[13] of 168 meV due to charge-quadrupole interactions.

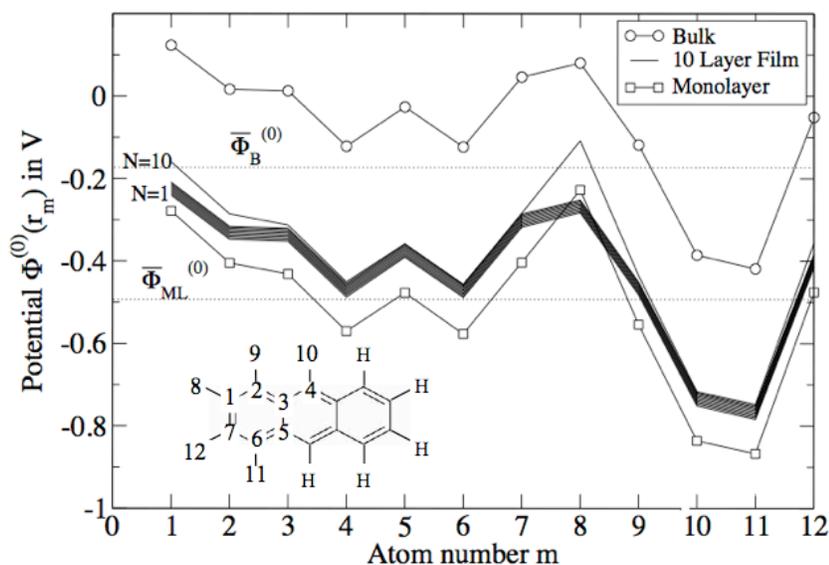

Figure 4. Gas-phase B3LYP potentials at anthracene atoms in a crystal, an *ab* monolayer and a film of 10 *ab* layers. Dashed lines are the average C-atom potential of the central ring.

# 4. Discussion

The self-consistent calculation of electronic polarization in crystalline thin films contains several inputs and approximations: the film's structure, taken from the crystal; no intermolecular overlap; discrete atomic charges and potentials; $\tilde{\alpha}$ to retain the molecular polarizability. Comparisons to bulk data, to microelectrostatic results or to films have been good.[14,4,5] The model reduces to induced atomic dipoles for point molecules with isotropic $\rho^{(0)}(\mathbf{r})$. Self-consistent treatments of atomic solid have been reported independently.[19,26,27] We regain microelectrostatic models[8,9] for molecules as polarizable points. Here we have focused on the depth and surface dependence of ionization in films of molecules with anisotropic $\rho^{(0)}(\mathbf{r})$.

There are several XPS studies of the binding energy (BE) of core holes in thin films[26] and clusters[27] of noble gases. BEs at the surface and interior of clusters of $10^3$-$10^5$ atoms are clearly resolved.[27,28] So are BE differences between a monolayer and a multilayer on a substrate.[26] Theoretical modeling[19] requires the inelastic mean free path $\lambda$ of photoelectrons as a function of photon energy h$\nu$ that governs the distribution of ionization sites. XPS linewidths in clusters or films of noble gases are about twice as narrow as in organic molecular films. The latter are broadened by molecular (vibrational) relaxation and by shape anisotropy. Nevertheless, weak van der Waals interactions are characteristic of both noble gas crystals and organic molecular crystals. Similar electronic polarization has consequently been widely assumed, until recently questioned.[6,29,30]

$P_+(n) + P_-(n)$ in Fig. 2 closely resembles Fig. 4 of ref. 4 aside from a small change of the bulk value for a more recent crystal structure and slightly different parameters. The 330 meV shift between $n = 9$ and 10 is the same. Casu *et al.*[6,29,30] misread 330 meV as referring to $P_+$ instead of $P_+ + P_-$. They studied C(1s) core holes in 4 nm PTCDA films (N $\sim$ 10) and found negligible BE changes within an estimated 100 meV resolution. Their data and detailed simulations[6] that include $\lambda$ rule out large C(1s) BE shifts in PTCDA and two other films. However, the shape of PTCDA molecules reduces the 9-10 shift of $P_+$ to only 119 meV in Fig. 2 while increasing the $P_-$ shift to 215 meV. Moreover, since $P_+(n)$ is not monotonic for the (102) surface and photoelectrons from several layers contribute, the calculated[16] BE shift is <100 meV. Hence PTCDA is not favorable for measuring surface core level shifts. Anthracene's 9-10 shift of 112 meV and monotonic $P_+(n)$ in Fig. 3 gives larger BE shifts that are somewhat less than the early estimated difference[31] of 200 meV between surface and bulk. Accurate measurement and analysis of BE shifts at surfaces of organic molecular films remain to be done.

The calculated $P_+(n)$ so far are for valence holes. To model C(1s) or other core holes, we use the equivalent-core approximation[32,33] with an N atom replacing the C atom in the cation. The charge density $\rho_+^{(0)}(\mathbf{r})$ is somewhat different,[16] but it is still delocalized over the molecule as shown[3] for a C(1s) hole in $C_{60}$. The $C_{60}$ films in Fig. 3 have a 60 meV shift between $n = 9$ and 10. Maxwell *et al*.[3] reported comparable shifts between the first and second $C_{60}$ monolayer on Al(1,1,1), Al(1,1,0) and Au(1,1,0). They view the first layer as partly metallized, so that the metal is not inert and van der Waal interactions start at the second layer. They find[3] $P_+ = 0.86$ eV for $I_G - I(s)$ for an N = 4 film on Al(1,1,1) and discuss both experimental and theoretical uncertainties. While $P_+ = 0.86$ eV agrees well with the calculated mean value in Fig. 3 for N = 10, finer BE variations in layers are still open.

We began by noting that accurate BE data in crystalline organic films invite improved modeling of $P_+$ and that electronic polarization is important in general for localized charges in organic electronics. Molecular shape leads to significant and often overlooked differences between $P_+$ and $P_-$. The detection of surface core level shifts is most promising in films with large $P_+(n)$ variations, and such variations are related to both electrostatic potentials at surfaces and to charge redistribution. The self-consistent treatment of electronic polarization is applicable to valence or core holes at different crystal faces as well as in films.

**Acknowledgements**. ZGS thanks Eugene Tsiper for stimulating correspondence on electronic polarization. We gratefully acknowledge support for work by the National Science Foundation under the MRSEC program (DMR-0819860).